\documentclass[aps,prl,twocolumn,showpacs,nofootinbib,superscriptaddress,10pt,floatfix]{revtex4-2}

\usepackage{amsmath,amssymb}
\usepackage{graphicx}
\usepackage{amsfonts,bm,tensor}
\usepackage{fnpct}
\usepackage{comment}
\usepackage{ifpdf}
\usepackage{ascmac}
\usepackage{physics}
\usepackage{slashed}
\usepackage{color}
\usepackage[dvipsnames]{xcolor}
\usepackage[mathscr]{eucal}
\usepackage[utf8]{inputenc}
\usepackage{cancel}
\usepackage{subcaption}
\usepackage{soul}
\usepackage{simpler-wick}
\usepackage{booktabs}
\usepackage{array}

\usepackage{hyperref}
\hypersetup{%
	colorlinks = true,
	linkcolor  = MidnightBlue,
	urlcolor   = BrickRed,
	citecolor  = MidnightBlue
}%

\newcommand{\Epsilon}{\mathcal{E}}

\begin{document}
	\title{Astrophysical Graviton Squeezing Can Be Hidden in the Far-Field}
\author{Cheng-Jun Fang}
\email{fangchengjun@itp.ac.cn}
\affiliation{Institute of Theoretical Physics, Chinese Academy of Sciences (CAS), Beijing 100190, China}
\affiliation{School of Physical Sciences, University of Chinese Academy of Sciences, Beijing 100049, China}

\author{Zong-Kuan Guo}
\email{guozk@itp.ac.cn}
\affiliation{Institute of Theoretical Physics, Chinese Academy of Sciences (CAS), Beijing 100190, China}
\affiliation{School of Physical Sciences, University of Chinese Academy of Sciences, Beijing 100049, China}
\affiliation{School of Fundamental Physics and Mathematical Sciences, Hangzhou Institute for Advanced Study, University of Chinese Academy of Sciences, Hangzhou 310024, China}

\author{Zhen-Hong Lyu}
\email{lyuzhenhong@itp.ac.cn}
\affiliation{Institute of Theoretical Physics, Chinese Academy of Sciences (CAS), Beijing 100190, China}
\affiliation{School of Physical Sciences, University of Chinese Academy of Sciences, Beijing 100049, China}

\author{Jing Shu}
\email{jshu@pku.edu.cn}
\affiliation{School of Physics and State Key Laboratory of Nuclear Physics and Technology, Peking University, Beijing 100871, China}
\affiliation{Center for High Energy Physics, Peking University, Beijing 100871, China}
\affiliation{Beijing Laser Acceleration Innovation Center, Huairou, Beijing 101400, China}

\author{Yu-Heng Sun}
\email{sunyuheng@itp.ac.cn}
\affiliation{Institute of Theoretical Physics, Chinese Academy of Sciences (CAS), Beijing 100190, China}
\affiliation{School of Physical Sciences, University of Chinese Academy of Sciences, Beijing 100049, China}

\author{Zi-Zheng Zhou}
\email{zhouzizheng@itp.ac.cn}
\affiliation{Institute of Theoretical Physics, Chinese Academy of Sciences (CAS), Beijing 100190, China}
\affiliation{School of Physical Sciences, University of Chinese Academy of Sciences, Beijing 100049, China}

\begin{abstract}
While localized astrophysical sources can generate macroscopic graviton squeezing, their observable quantum signatures at far-field detectors remain unresolved. In this work, we investigate the propagation dynamics of the squeezed states using spatial quantum optics methods to evaluate correlation functions accessible to a local observer. Crucially, we reveal a severe kinematic conflict in same-cone measurements, which highly suppresses local quantum coherence. Consequently, these macroscopically squeezed states appear classically thermal to a single detector. Our results demonstrate that global squeezing does not guarantee local observability, and the measurable quantum signatures may be significantly weaker than what would be expected from the overall squeezing parameter of the state.
\end{abstract}
\maketitle
%%%%%%%%%%%%%%%%%%%%%%%%%%%%%%%%%%%%%
\noindent\textbf{\emph{Introduction and summary.}}\label{sec:intro}
%%%%%%%%%%%%%%%%%%%%%%%%%%%%%%%%%%%%%
The quantum nature of gravity remains one of the most profound unresolved
questions in fundamental physics~\cite{Polchinski_1998,Bianconi:2024aju}.
Although individual gravitons are expected to be extraordinarily difficult to
detect~\cite{Dyson:2013hbl,Tobar:2023ksi,Carney:2023nzz}, macroscopic quantum states of the gravitational field can carry statistical signatures accessible to existing and near-future detectors~\cite{Parikh:2020kfh,Parikh:2020nrd,Parikh:2020fhy,Kanno:2021gpt}.
% may reveal themselves through statistical signatures rather than single-particle events~\cite{Parikh:2020kfh,Parikh:2020nrd,Parikh:2020fhy,Kanno:2021gpt}.
Highly squeezed graviton states are therefore natural targets for gravitational-wave (GW) observations~\cite{Kanno:2020usf,Ikeda:2025uae, Hertzberg:2021rbl}: their large occupation numbers make them detectable, while their non-classical correlations distinguish them from thermal noise.
% as they can have large occupation numbers while retaining non-classical correlations.

This prospect has spurred a search for concrete production mechanisms. Inflation naturally generates squeezed primordial gravitons~\cite{Grishchuk:1989ss,Grishchuk:1990bj, Albrecht:1992kf,Bianchi:2024jmn}, but cosmological evolution, decoherence, and detector limitations make extracting their quantum signatures exceedingly difficult~\cite{Burgess:2006jn,Kiefer:2008ku,Micheli:2022tld,Ning:2023ybc,Takeda:2025cye,Sano:2025ird,Martin:2015qta}. Consequently, late-time astrophysical phenomena have attracted growing attention as alternative sources of macroscopic graviton squeezing~\cite{Dorlis:2025amf,Dorlis:2025zzz,Kanno:2025how,Kanno:2025fpz,Dorlis:2026gth,Guerreiro:2025mcu,Guerreiro:2025sge,Manikandan:2025dea}.

% Motivated by the observational capabilities, recent theoretical attention has focused on finding possible sources. While inflation is well known to generate squeezed primordial gravitons~\cite{Grishchuk:1989ss,Grishchuk:1990bj, Albrecht:1992kf,Bianchi:2024jmn}, extracting quantum signatures from these relic states is expected to \lzh{be highly challenging due to} cosmological evolution, decoherence, and detector limitations~\cite{Burgess:2006jn,Kiefer:2008ku,Micheli:2022tld,Ning:2023ybc,Takeda:2025cye,Sano:2025ird,Martin:2015qta}. Consequently, late-time astrophysical phenomena have emerged as highly promising alternative candidates for generating macroscopic quantum states of gravitons~\cite{Dorlis:2025amf,Dorlis:2025zzz,Kanno:2025how,Kanno:2025fpz,Dorlis:2026gth,Guerreiro:2025mcu,Guerreiro:2025sge}.

%Despite the growing interest in these astrophysical sources, a critical gap remains. The precise relation between the intrinsic squeezing generated globally at the source and the actual quantum coherence measured by a realistic far-field detector has not been rigorously established. This gap fundamentally stems from the localized nature of astrophysical sources. To correctly evaluate the observable quantum signatures, one must account for the local spatial properties of the generated entangled states, which naturally requires extending the formalism of spatial quantum optics~\cite{Kolobov:1999zz,PhysRevA.50.3349,Lugiato:2002mbh, PhysRevA.57.3123,Pittman:1995hxu} to the gravitational regime.

Yet a critical question has been overlooked: does the squeezing of the global momentum-space state survive as measurable quantum coherence in a local far-field detector? Unlike cosmological particle production, which fills all of space homogeneously, a localized source produces entangled states with inherently
local spatial structure. Correctly evaluating the observable quantum signatures therefore requires extending the formalism of spatial quantum optics~\cite{Kolobov:1999zz,PhysRevA.50.3349,Lugiato:2002mbh,
PhysRevA.57.3123,Pittman:1995hxu} to the gravitational regime.

We carry out this extension by formulating an effective field theory (EFT) description of squeezed graviton-state generation, which systematically captures the properties of the source without relying on a specific model~\cite{Donoghue:1994dn,Das:2025kyn}. A broad class of quadratic interactions relevant for parametric down-conversion~\cite{PhysRevA.31.2409,PhysRevLett.57.2520} and four-wave mixing~\cite{Garay-Palmett:07,PhysRevA.72.033801} processes can be written schematically as
% In this Letter, we study squeezed graviton states from the viewpoint of effective field theory~\cite{Donoghue:1994dn,Das:2025kyn} to avoid restricting on specific models. A broad class of quadratic interactions
% relevant for parametric down-conversion~\cite{PhysRevA.31.2409,
% PhysRevLett.57.2520} and four-wave mixing~\cite{Garay-Palmett:07,
% PhysRevA.72.033801} processes can be written schematically as
	\begin{equation}
		\hat{H}_{\rm int}(t) = \int d^3x \, \xi^{ijmn}_{\mu\nu\cdots}({\mathbf{x}},t) \partial^{\mu\cdots}\hat h_{ij} \partial^{\nu\cdots}\hat h_{mn}\,.
	\end{equation}
Here $i,j$ and $m,n$ are the tensor indices of the two graviton fields. The interaction can contain either no derivatives or derivatives of various orders acting on $h_{ij}$, with $\mu,\nu\dots$ denoting the additional indices introduced by these derivatives. All source-specific information, including the coupling constants and contraction structure, is absorbed into the classical tensor field $\xi$. A far-field detector, however, does not directly probe this global state; it is sensitive to local coherence properties of the radiated field.

This distinction is familiar in spatial quantum optics~\cite{Kolobov:1999zz,PhysRevA.50.3349,Lugiato:2002mbh,
PhysRevA.57.3123,Pittman:1995hxu,Giovannini:2010xg}. There, local quantum signatures are characterized not only by particle number operators, but by coherence functions built from the local energy flux of the electric field. For gravitational radiation, the local energy flux is governed by $\partial_t \hat h_{ij}\partial_t \hat h_{ij}$~\cite{Isaacson:1968zza}; the gravitational analog of the electric field amplitude is therefore the projected strain rate $\hat E^{(\pm)}(x)\equiv\partial_t\hat h_D^{(\pm)}(x)$, where $x$ denotes a spacetime point. Here $\hat h_D$ denotes the detector-projected gravitational strain to be defined below, and $(\pm)$ denotes its positive- and negative-frequency parts. We define the normally ordered coherence function
\begin{equation}
g^{(2)}(x,x)
=
\frac{
\left\langle
\hat E^{(-)}(x)\hat E^{(-)}(x)
\hat E^{(+)}(x)\hat E^{(+)}(x)
\right\rangle
}{
\left\langle
\hat E^{(-)}(x)\hat E^{(+)}(x)
\right\rangle^2
}\,.
\end{equation}
% For gravitational radiation, we identify $\hat E^{(\pm)}(x)\equiv\partial_t\hat h_D^{(\pm)}(x)$, where $\hat h_D\equiv \Pi_D^{ij}\hat h_{ij}$ is the detector-projected gravitational strain --- the gravitational analogue of the local electric field in Glauber's coherence theory --- and $(\pm)$ denotes its positive- and negative-frequency parts~\cite{Isaacson:1968zza}. Here $D$ denotes the detector-projected radiative strain channel, with the explicit projection specified below. The normally ordered second-order degree of coherence is then
According to Glauber's quantum theory of measurement~\cite{PhysRevLett.10.84,PhysRev.130.2529}, $g^{(2)}$ is directly related to coincidence rates of absorption-based detectors. In principle, these statistics can be inferred from the quantum noise of
continuous strain in laser
interferometers~\cite{Amelino-Camelia:1998mjq,Amelino-Camelia:1999vks,
Guerreiro:2019vbq,Parikh:2020kfh,Coradeschi:2021szx} or probed more directly
with resonant cavity graviton detectors~\cite{Cai:2025fpe,Chen:1994ch,Coradeschi:2021szx}
through Hanbury--Brown--Twiss (HBT)
experiments~\cite{Brown:1956zza,Kanno:2019gqw,Kanno:2018cuk,Kanno:2024gjt}.

In this Letter, we show that such local far-field coherence can be parametrically much weaker than the global squeezing of the out-state: a macroscopically squeezed graviton state, therefore, can appear thermal to a single far-field observer.
For a smooth localized source whose pair-production envelope has a total-momentum width  $k_*$
much smaller than the typical graviton energy $\Epsilon$, the leading-order far-field expansion yields
\begin{equation}
g^{(2)}-2 \sim \mathcal{O}\left[ \exp\left(-\frac{\Epsilon}{k_*}\right) \right]\,.
\end{equation}
Thus, source-level graviton squeezing and local far-field observability are not equivalent.

%%%%%%%%%%%%%%%%%%%%%%%%%%%%%%%%%%%%%
\noindent\textbf{\emph{Squeezed gravitons from local sources.}}\label{sec:squeezed}
%%%%%%%%%%%%%%%%%%%%%%%%%%%%%%%%%%%%%
Starting from the EFT interaction, we expand $\hat h_{ij}$ in mode operators and apply the rotating-wave approximation~\cite{Scully_Zubairy_1997}. The resulting interaction-picture Hamiltonian captures the momentum-space structure of the squeezed state produced by the source and takes the form
\begin{equation}
\hat H_{\rm int}(t)
=
\frac12\int dI\,dJ\,
\left[
\mathcal F_{IJ}(t)\hat a_I^\dagger\hat a_J^\dagger
+
\text{h.c.}
\right]\,,
\end{equation}
where $I=(\mathbf p,s)$ and $J=(\mathbf q,s')$ are momentum--polarization
indices, $s,s'$ denote graviton tensor polarizations,
$\int dI\equiv\sum_s\int d^3p$, and $\mathcal F_{IJ}=\mathcal F_{JI}$.
Writing $p\equiv|\mathbf p|$ and $q\equiv|\mathbf q|$, we factor out the
fast oscillation and define the slowly varying envelope $\Xi_{IJ}(t)$ via
$\mathcal F_{IJ}(t)\equiv \Xi_{IJ}(t)\,e^{i(p+q)t}$, with
\begin{equation}
\Xi_{IJ}(t)\equiv
\frac{
\mathcal P^{ss'}_{ijmn}{}^{\mu\cdots\nu\cdots}(\mathbf p,\mathbf q)}
{2\sqrt{pq}}\,
\widetilde\xi^{ijmn}_{\mu\cdots\nu\cdots}
(\mathbf p+\mathbf q,t)\,.
\end{equation}
The factor
$\mathcal P^{ss'}_{ijmn}{}^{\mu\cdots\nu\cdots}$ comes from the derivatives
acting on the two graviton fields and their polarization tensors, while
$\widetilde\xi^{ijmn}_{\mu\cdots\nu\cdots}(\mathbf K,t)
\equiv\int d^3x\,e^{-i\mathbf K\cdot\mathbf{x}}
\xi^{ijmn}_{\mu\cdots\nu\cdots}(\mathbf{x},t)$; further details are given in
Appendix~\ref{app:squeezing}.

The quadratic Hamiltonian maps the in-operators into time-dependent
Bogoliubov-transformed operators, $\hat b_I(t) =\int dJ\, \left[ \alpha_{IJ}(t)\hat a_J + \beta_{IJ}(t)\hat a_J^\dagger \right]$, and the evolution of the coefficients follows directly from the Heisenberg equation and can be written in matrix form as $\dot{\mathsf S}=\mathsf K\mathsf S$,
with
$\mathsf S(-\infty)=\mathbb I$, where
	\begin{equation}
		\mathsf S\equiv
		\begin{pmatrix}
			\alpha(t)&\beta(t)\\
			\beta^*(t)&\alpha^*(t)
		\end{pmatrix},
		\qquad
		\mathsf K\equiv
		\begin{pmatrix}
			0&-i\mathcal F(t)\\
			i\mathcal F^*(t)&0
		\end{pmatrix}\,.
	\end{equation}
Keeping the first term in the Magnus expansion, this time evolution
corresponds to a multi-mode squeezed-state generating operator~\cite{PhysRevA.47.733,Blanes:2008xlr,Quesada:2014kmr}.
The formal solution is
$\mathsf S(t)=\mathbb I+\int_{-\infty}^{t}dt_1\, \mathsf K(t_1)\mathsf S(t_1)$.
{For simplicity, we define
	\begin{equation}
		\begin{pmatrix}
			Q_{IP}(t)\\
			{R_{IP}(t)}
		\end{pmatrix}
		\equiv
		\int dJ\, \Xi_{IJ}(t)e^{iqt}
		\begin{pmatrix}
			\alpha^*_{JP}(t)\\
			{\beta^*_{JP}(t)}
		\end{pmatrix}\,.
	\end{equation}
Using its Fourier transform over $t$, the solution $\mathsf S$ gives
	\begin{equation}
		\begin{pmatrix}
			\beta_{IP}(t)\\
			{\alpha_{{\rm sc},IP}(t)}
		\end{pmatrix}
		=
		\int\frac{d\omega}{2\pi}\,
		\frac{e^{-i(\omega-p)t}}{\omega-p+i0}
		\begin{pmatrix}
			\widetilde Q_{IP}(\omega)\\
			\widetilde R_{IP}(\omega)
		\end{pmatrix}\,,
	\end{equation}
where $\alpha_{{\rm sc},IP}\equiv\alpha_{IP}-\delta_{IP}$ is the scattered coefficient.}  These Bogoliubov coefficients, $\alpha_{\rm sc}$ and $\beta$, fully characterize the momentum-space structure of the generated squeezed state. Crucially, the upper limit of the time integral in $\mathsf S$ produces the retarded pole prescription $\omega-p+i0$ for these coefficients rather than an ordinary Fourier transform, a causal structure that ultimately dictates the far-field radiation.

Since astrophysical sources are localized, we examine how real-space locality of the source appears in
the momentum-space coefficients of this multi-mode squeezed state. 
We consider a generic localized profile $f(\mathbf{x})$ and
$\tilde{f}_{\mathbf K}=\int d^3x\,f(\mathbf{x})e^{-i\mathbf K\cdot\mathbf{x}}$.
Then
$|\nabla_{\mathbf K}\tilde{f}_{\mathbf K}| \le\int d^3x\,|\mathbf{x}|\,|f(\mathbf{x})| \equiv R_{\rm eff}^{(1)}\|f\|$.
Higher momentum derivatives obey analogous bounds controlled by higher
spatial moments. Thus, a localized source produces a smooth momentum-space envelope whose variation scale is set by the inverse source size.\footnote{This contrasts with cosmological sources~\cite{Grishchuk:1989ss,Grishchuk:1990bj}, where spatial homogeneity enforces momentum conservation through a delta function.}.

In the source envelope $\Xi$, the source Fourier variable is
$\mathbf K=\mathbf p+\mathbf q$.  Therefore, $\Xi$ inherits the same
momentum smoothness.  Derivatives of the kinematic polynomial
$\mathcal P_{IJ}$ and of the mode normalization
only introduce corrections on the scale of the gravitational wavelength.
Schematically,
	\begin{equation}
		|\nabla_{\mathbf p}\Xi_{IJ}| \lesssim L_\Xi\,\Xi_*\,, \qquad L_\Xi\sim R_{\rm eff}+p^{-1}\,,
	\end{equation}
where $\Xi_*$ is the majorant for $\Xi$. The precise smoothness estimates used
below are given in Appendix~\ref{app:locality}.

%%%%%%%%%%%%%%%%%%%%%%%%%%%%%%%%%%%%%
\noindent\textbf{\emph{Far-field expansion.}}\label{farfield}
%%%%%%%%%%%%%%%%%%%%%%%%%%%%%%%%%%%%%
Consider a far-field detector at $\mathbf{x}=r\hat{n}$,
with a smooth frequency window $W_D(\omega)$ of finite
bandwidth~\cite{LIGOScientific:2003jxj,Allen:1997ad}.  Defining
$\mathcal C_{I}\equiv -ip\,\Pi_D^{ij}\epsilon^{(s)}_{ij}(\mathbf p)/\sqrt{2p}$,
where $\Pi_D^{ij}$ is the detector projection tensor and $\epsilon^{(s)}_{ij}(\mathbf p)$ is the graviton polarization tensor, the positive-frequency part of the projected strain rate is
	\begin{equation}
		\partial_t\hat h_D^{(+)}(x) = \int dI\, W_D(p)\mathcal C_{I} e^{i(\mathbf p\cdot\mathbf{x}-pt)} \hat b_{I}\,.
	\end{equation}
The negative-frequency part follows by Hermitian conjugation, so we display only the positive part in what follows.
The out-operator $\hat b_I$ decomposes into the direct freely propagating term $\hat a_I$ and the scattering part $\hat b_{I,\rm sc}=\int dJ\, \left[ \alpha_{{\rm sc},IJ}(t)\hat a_J + \beta_{IJ}(t)\hat a_J^\dagger \right]$.  Substituting this into the strain rate yields
\begin{equation}
\label{sc}
\partial_t\hat h_{D,{\rm sc}}^{(+)}
\!=\!
\int\!dJ\frac{d\omega}{2\pi}\,
e^{-i\omega t}
\left[
\mathcal I_{J}^{(Q)}(\omega)\hat a_J^\dagger
+
(Q\!\leftrightarrow\!R)
\right]\,.
\end{equation}

Here $\mathcal I_{J}^{(Q)}$ denotes the integral over the observed mode index $I$. The far-field expansion amounts to simplifying this observed-mode integral. The notation $(Q\leftrightarrow R)$ denotes the analogous term obtained by replacing $Q\!\rightarrow\! R$ and $\hat a_J^\dagger\!\rightarrow\! \hat a_J$.
For the $Q$-part, we define $A_{IJ}^{(Q)}(\omega)\equiv W_D(p)\mathcal C_I\widetilde Q_{IJ}(\omega)$ and write
\begin{equation}
\begin{aligned}
\mathcal I_{J}^{(Q)}(\omega;x)
&\equiv
\int dI\,
\frac{
e^{i\mathbf p\cdot\mathbf x}
}{
\omega-p+i0
}
A_{IJ}^{(Q)}(\omega)
\\ 
&=I^{(Q)}_+ + I^{(Q)}_- + I^{(Q)}_{\rm rem}\,.
\end{aligned}
\end{equation}
Here $I^{(Q)}_\pm$ and $I^{(Q)}_{\rm rem}$ follow from the angular integral over $\Omega_{\hat p}$; the $R$-part admits the same decomposition. The angular phase is stationary at $\hat p=\pm\hat n$, so $I_\pm$ capture the geometric endpoint contributions of the two branches, while $I_{\rm rem}$ contain the remaining angular corrections.
For arbitrary $J$ and either $Q$ or $R$,
\begin{equation}
I_\pm
=
\pm
\int_{\mathbb R}dp\,
\frac{2\pi p}{ir}\,
A(\omega;\pm p\hat n)
\frac{e^{\pm ipr}}{\omega-p+i0}\,.
\end{equation}
The remainder $I_{\rm rem}$ is suppressed by the rapidly oscillating phase in the integral. Under the far-field condition $r\gg pL_\Xi^2$, source locality ensures that the amplitude $A$ is sufficiently smooth, so the remainder is parametrically smaller than the leading endpoint terms. The details of this angular endpoint expansion are given in Appendix~\ref{app:angular}.

\begin{figure}[t]
	\centering
	\includegraphics[width=\columnwidth]{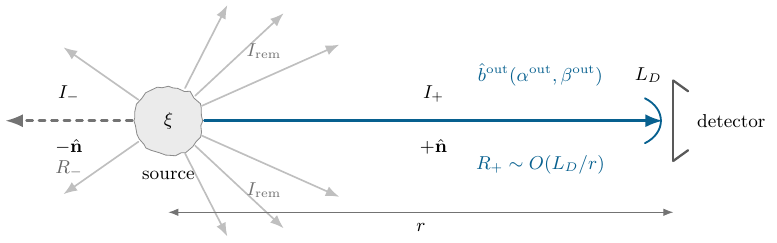}
	\caption{Far-field projection onto the outgoing branch $I_+$.}
	\label{fig:far-field-geometry}
\end{figure}

To identify which geometric branch produces the far-field radiation, we now evaluate the radial integrals $I_\pm$. Both are determined by the identity
$({\omega-p+i0})^{-1} = {\rm PV}(\omega-p)^{-1} -i\pi\delta(\omega-p)$: 
\begin{equation}
    \begin{aligned}
        I_+ &= -2\pi i\,\frac{2\pi\omega}{ir}\!{A(\omega;\omega\hat{n})}e^{i\omega r}+{R_+(\omega;r)}\,,\\
        I_- &=  {R_-(\omega;r)}\,.
    \end{aligned}
\end{equation}
Combined with the overall time factor $e^{-i\omega t}$, $I_+$ yields the retarded phase $e^{-i\omega(t-r)}$. For $I_-$, the principal-value and delta-function contributions exactly cancel, leaving no advanced radiation term proportional to $e^{-i\omega(t+r)}$. The remainders $R_\pm$ consist of smooth amplitudes multiplied by rapidly oscillating phases; integration by parts therefore gives
	\begin{equation}
		|{R_\pm(\omega;r)}| \sim \mathcal{O}\left(\frac{L_D}{r}\right)\,,
	\end{equation}
where $L_D\sim L_\Xi+\omega_D^{-1}$ is the radial smoothness scale including the detector bandwidth $\omega_D$. Thus $R_+$ and $R_-$ are remainders that contain only near-field effects. The details of full pole extraction are given in Appendix~\ref{app:retarded-pole}. The resulting source-to-detector geometry is summarized in Fig.~\ref{fig:far-field-geometry}.

The local field, therefore, expands into retarded outgoing operators in the far field,
	\begin{equation}\label{far E}
		\begin{aligned}
			&\hat E^{(+)}(x)
			=
			\partial_t\hat h_{D}^{(+)}(x)
			\\
			&\simeq
			\sum_s\int_0^\infty d\omega\,
			\frac{2\pi\omega}{ir}
			W_D(\omega)\mathcal C_{\omega\hat{n},s}
			e^{-i\omega(t-r)}
			\hat b^{{\rm out}}_{\omega\hat{n},s}\,,
		\end{aligned}
	\end{equation}
where $\hat b^{\rm out}_{\omega\hat{n},s} = \hat a_{\omega\hat{n},s} -i\int dJ\left[ \widetilde R^s_{\omega\hat{n},J}(\omega)\hat a_J + \widetilde Q^s_{\omega\hat{n},J}(\omega)\hat a_J^\dagger \right]$.

The physical implication of $\widetilde R$ and $\widetilde Q$ becomes clear when they are related to the future asymptotic Bogoliubov coefficients $\alpha^{\rm out}$ and $\beta^{\rm out}$. Direct calculation from their definitions yields $\beta^{\rm out}_{IJ} = -i\widetilde Q^s_{\mathbf{p},J}(p)$ and $\alpha^{\rm out}_{{\rm sc},IJ} = -i\widetilde R^s_{\mathbf{p},J}(p)$. This reveals that a far-field observer effectively measures the asymptotic out-state. In the leading-Magnus approximation, these out-coefficients are generated by the multi-mode squeezing matrix $\mathcal G_{IJ} =-i\int_{-\infty}^{+\infty}dt\,\mathcal F_{IJ}(t)$~\cite{PhysRevA.47.733} via
\begin{equation}
		\label{eq:magnus-out} \alpha^{\rm out} = \cosh\sqrt{\mathcal G\mathcal G^\dagger}\,, \qquad \beta^{\rm out} = \mathcal G \frac{\sinh\sqrt{\mathcal G^\dagger\mathcal G}} {\sqrt{\mathcal G^\dagger\mathcal G}}\,.
\end{equation}
	
%%%%%%%%%%%%%%%%%%%%%%%%%%%%%%%%%%%%%
\noindent\textbf{\emph{Same-cone suppression.}}\label{samecone}
%%%%%%%%%%%%%%%%%%%%%%%%%%%%%%%%%%%%%
For a far-field observer measuring retarded radiation, the angular aperture
is negligible compared with the source distance.  We therefore evaluate the
zero-delay same-cone observable at $x_a=x_b=x$, using the retarded
detector channel $\hat E^{(+)}(x)$ in Eq.~\eqref{far E}. We define its normal and anomalous moments as
$N_D = \langle \hat E^{(-)}\hat E^{(+)}\rangle$, $M_D = \langle \hat E^{(+)}\hat E^{(+)}\rangle.$ Here $N_D$ is the received energy intensity in the detector channel, while
$M_D$ is the anomalous coherence. For the same-cone estimate, they can be schematically written in terms of overlap integrals,
\begin{equation}
	\label{eq:same-cone-NM}
	{N_D \sim \int dJ\,\beta^{{\rm out}*}_{IJ}\beta^{\rm out}_{I^\prime J}\,, \qquad
	M_D \sim \int dJ\,\alpha^{\rm out}_{IJ}\beta^{\rm out}_{I^\prime J}}\,.
\end{equation}
Here, $I$ and $I^\prime=(\mathbf p^\prime,s^\prime)$ denote the detected modes whose momenta are both directed along the observer's line of sight $\hat{n}$, thus $\mathbf p^\prime \parallel \mathbf p \parallel \hat{n}$.
Since a multi-mode squeezed state is a Gaussian state, Wick's theorem gives
\begin{equation}
    g^{(2)} = 2+\frac{|M_D|^2}{N_D^2}\,.
\end{equation}
The corresponding two-point formula for a small angle aperture is given in
Appendix~\ref{app:same-cone}.

Because a single far-field observer is restricted to measuring same-direction graviton coherence, the quantum signatures of a broad class of multi-mode squeezed states can be observationally hidden. For astrophysical sources, the squeezing matrix $\mathcal G$ typically peaks near the anti-diagonal configurations $\vert{}\mathbf p+\mathbf q\vert{}\ll p+q$. This kinematic regime arises when the frequencies of the source are much larger than its characteristic momenta. Physically it describes non-relativistic sources with rest mass~\cite{Dorlis:2025amf}. We quantify this anti-diagonal behavior using a general exponential envelope: \begin{equation}
		\label{kernel} {|\mathcal G_{\mathbf p,\mathbf q}^{ss'}| \le G_0\exp\left(-c\frac{|\mathbf p+\mathbf q|}{k_*}\right)}\,.
	\end{equation}
Here $k_*$ is the total-momentum spread of the squeezing matrix, with {$k_*\ll p+q$} in the non-relativistic regime, and $c$ is a positive constant. Combined with the spatial locality discussed earlier, the momentum-space distribution of the source is constrained from both sides. The finite size of the source prevents this distribution from becoming arbitrarily narrow, while the non-relativistic kinematic condition restricts the available total momentum and prevents it from becoming excessively broad.
	
According to Eq.~\eqref{eq:magnus-out}, the Bogoliubov coefficients inherit the exponential envelope of the source. As a result, the amplitude of $\beta^{\rm out}$ peaks near the anti-diagonal $\mathbf q\simeq-\mathbf p$, whereas $\alpha_{\rm sc}^{\rm out}$ peaks near the diagonal $\mathbf q\simeq\mathbf p$. The proof is detailed in Appendix~\ref{app:same-cone}.

\begin{figure}[t]
\centering
\includegraphics[width=0.78\columnwidth]{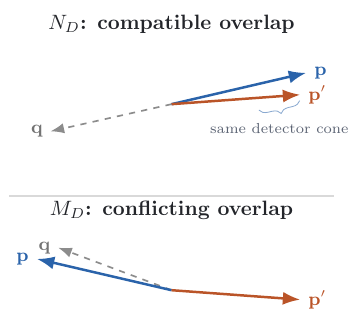}
\caption{Same-cone geometry: $N_D$ selects $\mathbf p\simeq\mathbf p'\simeq-\mathbf q$, while $M_D$ selects $\mathbf p\simeq\mathbf q\simeq-\mathbf p'$.}
\label{fig:same-cone-geometry}
\end{figure}

As illustrated in Fig.~\ref{fig:same-cone-geometry}, the anomalous same-cone amplitude $M_D$ in Eq.~\eqref{eq:same-cone-NM} is governed by the $\alpha$-$\beta$ overlap integral over the shared internal mode $J$. This integration requires the same internal momentum $\mathbf{q}$ to be simultaneously close to $\mathbf{p}$ and close to $-\mathbf{p}^\prime$. For same-cone observations $\mathbf{p}\simeq\mathbf{p}^\prime$, these two requirements are kinematically conflicting, suppressing the overlap via the large total momentum exponential tail. By contrast, the $\beta$-$\beta$ overlap in $N_D$ only requires $\mathbf{q}\simeq-\mathbf{p}$ and $\mathbf{q}\simeq-\mathbf{p}^\prime$, which are perfectly compatible when $\mathbf{p}\simeq\mathbf{p}^\prime$. Consequently, the ratio $M_D/N_D$ acquires a severe relative suppression factor $\eta\sim\exp(-d\Epsilon/k_*)$, where $\Epsilon$ is the characteristic same-cone energy scale in the detector band, and $d<c$ is a positive constant determined by the finite convolution estimates.

For a compact estimate, we map the detected energy flux to an effective squeezing parameter $r_{\rm eff}$ via $N_D\sim\sinh^2 r_{\rm eff}$. Using the decomposition $\alpha=\delta+\alpha_{\rm sc}$, the identity piece $\delta$ gives a fractional coherence of $1/\sinh r_{\rm eff}$, and the rescattered piece $\alpha_{\rm sc}$ gives $\tanh(r_{\rm eff}/2)$. Factoring in the kinematic suppression, we obtain
\begin{equation}
\frac{|M_D|}{N_D} \lesssim \exp\left(-{d}\frac{\Epsilon}{k_*}\right) \left[ \frac{1}{\sinh r_{\rm eff}} + \tanh\frac{r_{\rm eff}}{2} \right]\,.
\end{equation}
A full multimode distribution would only modify the order-one prefactors inside the bracket. Although the first term diverges as $r_{\rm eff}\to0$, this merely reflects the low-occupation limit: the normalized coherence artificially blows up, but the absolute signal $\vert{}M_D\vert{}^2\propto \sinh^2r_{\rm eff}$ vanishes, yielding no observable macroscopic excess.

For a populated squeezed channel $r_{\rm eff}\gtrsim 1$, the $r_{\rm eff}$-dependent factor in the estimate evaluates to $\mathcal{O}(1)$. Restoring the relative projection corrections from the earlier far-field expansion, the deviation from the chaotic HBT value becomes
\begin{equation}
g^{(2)}-2 = \mathcal{O}\left[ \exp\left(-2d\frac{\Epsilon}{k_*}\right) \right] + \mathcal{O}\left(\frac{L_D}{r}\right)\,.
\end{equation}
Since a purely thermal state yields exactly $g^{(2)}=2$, this residual suppression effectively hides the quantum signatures of a macroscopically squeezed state. The GWs appear thermal to a single far-field observer. A large global squeezing source alone is therefore insufficient for local quantum detection.

%%%%%%%%%%%%%%%%%%%%%%%%%%%%%%%%%%%%%
\noindent\textbf{\emph{Discussion.}}\label{sec:discussion}
%%%%%%%%%%%%%%%%%%%%%%%%%%%%%%%%%%%%%
The preceding analysis establishes a conditional no-go theorem for observing macroscopic graviton squeezing. Such a state will conceal its quantum correlations and appear classically thermal provided three physical conditions hold: (i) the source is non-relativistic, emitting entangled gravitons primarily as back-to-back pairs; (ii) the detector is localized, sampling only a single far-field line of sight; and (iii) the radiation propagates through near flat spacetime.

% \fcj{For a nonrelativistic axion cloud, this naturally translates to a nearly back-to-back emission, because the cloud carries negligible spatial momentum compared to its rest energy~\cite{Brito:2015oca}. However, this condition can fail for other kinds of sources. This breakdown occurs when the wavelength of the produced gravitons is larger than, or comparable to, the characteristic spatial variation scale of the source, meaning same-direction graviton pairs are no longer exponentially rare. Possible examples include relativistic axion clouds and graviton squeezing generated during black-hole orbital motion~\cite{Kanno:2025how,Kanno:2025fpz}. Future work should determine
% the angular distribution of the produced graviton pairs more completely. Furthermore, the large-momentum behavior of the squeezing kernel requires careful consideration in future works since it is sensitive to the small-scale real-space structure of the source.} 

Violating the first condition requires that the spatial momentum scale of the source matches the typical radiation energy ($k_* \sim \Epsilon$). This reveals the observation limitation of non-relativistic astrophysical sources like axion clouds~\cite{Dorlis:2025amf}, where annihilations enforce back-to-back emission and result in exponential suppression. If macroscopic squeezing occurs in highly dynamical environments like binary black hole systems~\cite{Kanno:2025how} 
or relativistic jets, collinear emission becomes kinematically possible, potentially reducing the suppression factor $\exp(-d\Epsilon/k_*)$ to order unity. Fully realizing this scenario requires future work to determine the complete angular distribution of the emitted pairs. Furthermore, the large-momentum behavior of the squeezing matrix must be carefully evaluated, since it is highly sensitive to the small-scale spatial structure of these relativistic sources.

% The second condition is the single-cone nature of the observable.  A
% back-to-back squeezed state still contains quantum correlations, but these correlations are
% shared between approximately opposite far-field directions. A detector
% network that accesses both partner cones would instead probe
% $g^{(2)}(x_a,x_b)$ with $x_a$ and $x_b$ in different directions.  This
% is unrealistic for distant astrophysical sources, but may become relevant in nearby settings. A concrete example is a solar axion halo, supported by the Sun's gravitational potential and potentially extending to Solar-System scales~\cite{Budker:2023sex}.

The second condition limits the observation to a single far-field cone. A back-to-back squeezed state inherently contains macroscopic quantum correlations, but these are distributed between approximately opposite directions. To recover the signal, a network of detectors accessing both partner cones must probe the cross-correlation $g^{(2)}(x_a,x_b)$ across different lines of sight. While coordinating such a baseline is unrealistic for distant astrophysical sources, it becomes physically relevant in nearby settings. A potential example is a solar axion halo~\cite{Budker:2023sex}, allowing local multi-detector cross-correlations.

Finally, the flat-propagation approximation breaks down near compact objects. Extremely strong curved-spacetime effects cause severe gravitational lensing. Consequently, graviton pairs initially emitted back-to-back in the local source frame can potentially have their trajectories deflected and redirected into the same far-field observation cone. This curved-spacetime propagation can therefore defeat the same-cone kinematic suppression, providing another astrophysical route to locally observable quantum correlations.

% Astrophysical entangled graviton sources open a new window onto the search
% for quantum gravitational effects.  To turn this possibility into observable
% signatures, one must understand how the produced quantum state is distributed
% across space.  Describing its spatial coherence in the language of spatial
% quantum optics is therefore an essential step toward identifying observable
% effects.

Astrophysical sources of entangled gravitons represent a highly promising frontier in the search for quantum gravitational effects. However, our results demonstrate that global momentum-space squeezing does not automatically guarantee local observability. To translate theoretical production mechanisms into realistic detection strategies, one must rigorously account for how these macroscopic quantum states are distributed across spacetime. Extending the formalism of spatial quantum optics to the gravitational sector is therefore an indispensable step toward unlocking the quantum signatures of gravity.

\medskip
\noindent\textit{Note added.} While this manuscript was about to be finished, a preprint~\cite{Miyauchi:2026cnt} appeared on arXiv that investigated a similar scenario. Although we independently reached consistent conclusions, our work relies on a different framework. 

\noindent\textit{Acknowledgments.}
We sincerely thank Panagiotis Dorlis for insightful discussions. Jing Shu is supported by the National Natural Science Foundation of China (No.12450006). Zong-Kuan Guo is supported by the National Natural Science Foundation of China (No. 12475067 and No. 12235019).

\bibliographystyle{apsrev4-1}
\bibliography{main}
\appendix

\section{Squeezing Hamiltonian and Bogoliubov Evolution}
\label{app:squeezing}

This appendix records the operator evolution used in the main text and
gives the tensor version of the squeezing matrix that is written
schematically there.  Writing the spacetime arguments explicitly as
$(\mathbf{x},t)$, we start from the general
quadratic interaction
	\begin{equation}
		\hat H_{\rm int}(t)=\int d^3x\,\xi^{ijmn}_{\mu\cdots\nu\cdots}(\mathbf{x},t)\partial^{\mu\cdots}\hat h_{ij}\partial^{\nu\cdots}\hat h_{mn}\,.
	\end{equation}
Here $\mu\cdots$ and $\nu\cdots$ are derivative multi-indices acting on
the first and second graviton fields, respectively, and repeated Lorentz
derivative indices are contracted.  The tensor susceptibility $\xi$ contains the
classical source profile, coupling constants, and derivative and tensor
contractions of the EFT operator.  As a concrete example, the axion-induced
graviton interaction
	\begin{equation}
		\mathcal L_I^{(2)} = -\frac{\kappa^2}{2}\, h_{im}h^m{}_{j}\,\partial^i b\,\partial^j b
	\end{equation}
contains no time derivatives of $h_{ij}$.  Treating $b(\mathbf{x},t)$ as a
classical source, its interaction Hamiltonian is therefore
	\begin{equation}
		\hat H_I^{(2)}(t) = \frac{\kappa^2}{2}\int d^3x\, \hat h_{im}\hat h^m{}_{j}\, \partial^i b\,\partial^j b \,.
	\end{equation}
This is a special case of the susceptibility form, with no derivatives
acting on the graviton fields and with
$\xi^{ijmn}(\mathbf{x},t)=(\kappa^2/2)\delta^{jm}\partial^i b(\mathbf{x},t)\partial^n b(\mathbf{x},t)$.

We use the positive-frequency convention
	\begin{equation}
		\hat h^{(+)}_{ij}(\mathbf{x},t)=\sum_s\int d^3p\,\frac{\epsilon^{(s)}_{ij}(\mathbf p)}{\sqrt{2p}}\,e^{i\mathbf p\cdot\mathbf{x}-ipt}\hat a_{\mathbf p,s}\,.
	\end{equation}
The negative-frequency part is $\hat h^{(-)}_{ij}(\mathbf{x},t)=[\hat h^{(+)}_{ij}(\mathbf{x},t)]^\dagger$ and supplies the creation operators in the pair-creation term.  For the displayed positive-frequency phase,
	\begin{equation}
		\partial^{\mu\cdots}\left[e^{i\mathbf p\cdot\mathbf{x}-ipt}\right]=D_+^{\mu\cdots}(\mathbf p)e^{i\mathbf p\cdot\mathbf{x}-ipt}\,.
	\end{equation}
The pair-creation formulas below use $D^{\mu\cdots}(\mathbf p)\equiv D_+^{\mu\cdots}(\mathbf p)^*$, the derivative eigenvalue for the Hermitian-conjugate phase.  We also define the spatial Fourier transform
	\begin{equation}
		\widetilde\xi^{ijmn}_{\mu\cdots\nu\cdots}(\mathbf K,t)\equiv\int d^3x\,e^{-i\mathbf K\cdot\mathbf{x}}\xi^{ijmn}_{\mu\cdots\nu\cdots}(\mathbf{x},t).
	\end{equation}
Keeping the slowly rotating $\hat a^\dagger\hat a^\dagger$ and
$\hat a\hat a$ terms gives
	\begin{equation}
		\hat H_I(t)=\frac12\int dI\,dJ\,\left[\mathcal F_{IJ}(t)\hat a_I^\dagger\hat a_J^\dagger+\mathcal F^*_{IJ}(t)\hat a_J\hat a_I\right].
	\end{equation}
where $I=(\mathbf p,s)$, $J=(\mathbf q,s')$, and
	\begin{equation}
		\mathcal F^{ss'}_{\mathbf p,\mathbf q}(t)
		=
		\Xi^{ss'}_{\mathbf p,\mathbf q}(t)
		e^{i(p+q)t}.
	\end{equation}
Only the part of $\mathcal F_{IJ}$ symmetric under $I\leftrightarrow J$
contributes to $\hat H_I$, because the antisymmetric part integrates to zero
against $\hat a_I^\dagger\hat a_J^\dagger=\hat a_J^\dagger\hat a_I^\dagger$
and similarly for the annihilation term.  Thus one may replace
$\mathcal F_{IJ}$ by its symmetric part when desired, without changing the
Hamiltonian.
The tensor envelope is
	\begin{equation}
		\Xi^{ss'}_{\mathbf p,\mathbf q}(t)=\frac{1}{2\sqrt{pq}}\,\widetilde\xi^{ijmn}_{\mu\cdots\nu\cdots}(\mathbf p+\mathbf q,t)\mathcal P^{ss'}_{ijmn}{}^{\mu\cdots\nu\cdots}(\mathbf p,\mathbf q),
	\end{equation}
with
	\begin{equation}
		\mathcal P^{ss'}_{ijmn}{}^{\mu\cdots\nu\cdots}(\mathbf p,\mathbf q)=D^{\mu\cdots}(\mathbf p)D^{\nu\cdots}(\mathbf q)\epsilon^{(s)*}_{ij}(\mathbf p)\epsilon^{(s')*}_{mn}(\mathbf q).
	\end{equation}
This is the precise tensor version of the schematic expression used in
the main text.  The source information enters through the Fourier transform of
the tensor susceptibility at total momentum $\mathbf K=\mathbf p+\mathbf q$,
while $\mathcal P^{ss'}_{ijmn}{}^{\mu\cdots\nu\cdots}$ contains the finite-order
derivative and polarization structure of the EFT interaction.

Let $i\partial_t\hat U(t,-\infty)=\hat H_I(t)\hat U(t,-\infty)$, and define
$\hat b_I(t)=\hat U^\dagger(t,-\infty)\hat a_I\hat U(t,-\infty)$.  The
only nontrivial commutator needed is
	\begin{equation}
		[\hat a_K^\dagger\hat a_L^\dagger,\hat a_I] = -\delta_{KI}\hat a_L^\dagger-\delta_{LI}\hat a_K^\dagger\,,
	\end{equation}
which gives
	\begin{align}
		\dot{\hat b}_I(t) \nonumber &= -i\int dJ\,\mathcal F_{IJ}(t)\hat b_J^\dagger(t), \\
		\dot{\hat b}_I^\dagger(t) &= i\int dJ\,\mathcal F^*_{IJ}(t)\hat b_J(t)\,.
	\end{align}
Expanding the evolved operator as
$\hat b_I(t)=\int dP[\alpha_{IP}(t)\hat a_P+\beta_{IP}(t)\hat a_P^\dagger]$
and comparing the coefficients of $\hat a_P$ and $\hat a_P^\dagger$
gives
	\begin{equation}
		\dot\alpha_{IP} = -i\int dJ\,\mathcal F_{IJ}\beta^*_{JP}, \qquad \dot\beta_{IP} = -i\int dJ\,\mathcal F_{IJ}\alpha^*_{JP}\,.
	\end{equation}
The retarded initial conditions are
$\alpha_{IP}(-\infty)=\delta_{IP}$ and $\beta_{IP}(-\infty)=0$.  In
matrix form this is
	\begin{equation}
		\dot{\mathsf S}(t)=\mathsf K(t)\mathsf S(t), \qquad \mathsf S(-\infty)=\mathsf I\,,
	\end{equation}
with $\mathsf S$ and $\mathsf K$ defined in the main text.  Equivalently,
	\begin{equation}
		\mathsf S(t) = \mathsf I+\int_{-\infty}^{t}dt_1\,\mathsf K(t_1)\mathsf S(t_1)\,,
	\end{equation}
which is the Volterra form used to fix the retarded pole prescription.

When the leading Magnus term is sufficient, the finite-time evolution is
already a multi-mode squeeze transformation,
	\begin{equation}
		\mathsf S_M(t)=
		\exp\!\begin{pmatrix} 0&\mathcal G(t)\\
		\mathcal G^*(t)&0 \end{pmatrix}\,.
	\end{equation}
where
	\begin{equation}
		\mathcal G_{IJ}(t) = -i\int_{-\infty}^{t}dt_1\,\mathcal F_{IJ}(t_1)\,.
	\end{equation}
Thus at any intermediate time the leading-Magnus coefficients are
	\begin{align}
		\alpha_M(t) \nonumber &= \cosh\sqrt{\mathcal G(t)\mathcal G^\dagger(t)}, \\
		\beta_M(t) &= \mathcal G(t) \frac{\sinh\sqrt{\mathcal G^\dagger(t)\mathcal G(t)}} {\sqrt{\mathcal G^\dagger(t)\mathcal G(t)}}\,.
	\end{align}
The out coefficients used in the main text are obtained by taking the same
form after the source has switched off, equivalently
$\mathcal G=\mathcal G(+\infty)$.  In this approximation the pair envelope
is directly inherited from the spacetime Fourier transform of the localized
source.

\section{Source Locality and Momentum Smoothness}
\label{app:locality}

This appendix justifies the momentum-smoothness estimates used in the
far-field analysis.  The point is simple: for a localized source whose
characteristic size is much smaller than the source-observer distance $r$,
its Fourier transform can oscillate in momentum space only with rates set by
the source size, and hence much more slowly than the propagation phase
$e^{ipr}$.

At each time $t$, differentiating the spatial Fourier transform defined
in Appendix~\ref{app:squeezing} brings down powers of the source position.
Thus, for any spatial multi-index $\alpha$,
	\begin{equation}
		\partial_{\mathbf K}^{\alpha}\widetilde\xi^{ijmn}_{\mu\cdots\nu\cdots}(\mathbf K,t)=\int d^3x\,(-i\mathbf{x})^\alpha\xi^{ijmn}_{\mu\cdots\nu\cdots}(\mathbf{x},t)e^{-i\mathbf K\cdot\mathbf{x}}.
	\end{equation}
For each component define the positive majorant and, for
$|\alpha|>0$, its spatial moment scale by
	\begin{equation}
		\xi^{ijmn}_{*,\mu\cdots\nu\cdots}(t)\equiv\int d^3x\,\left|\xi^{ijmn}_{\mu\cdots\nu\cdots}(\mathbf{x},t)\right|,
	\end{equation}
	\begin{equation}
		\left(R^{ijmn}_{\mu\cdots\nu\cdots;\alpha}(t)\right)^{|\alpha|}\equiv\frac{\int d^3x\,|\mathbf{x}^\alpha|\left|\xi^{ijmn}_{\mu\cdots\nu\cdots}(\mathbf{x},t)\right|}{\xi^{ijmn}_{*,\mu\cdots\nu\cdots}(t)},
	\end{equation}
whenever the denominator is nonzero.  Finiteness of these componentwise
moments gives the following bound, with the zeroth-order convention
$\left(R^{ijmn}_{\mu\cdots\nu\cdots;0}(t)\right)^0=1$:
	\begin{equation}
		\left|\partial_{\mathbf K}^{\alpha}\widetilde\xi^{ijmn}_{\mu\cdots\nu\cdots}(\mathbf K,t)\right|\le \left(R^{ijmn}_{\mu\cdots\nu\cdots;\alpha}(t)\right)^{|\alpha|}\xi^{ijmn}_{*,\mu\cdots\nu\cdots}(t).
	\end{equation}
We now apply the product rule to the tensor expression for $\Xi$ derived
in Appendix~\ref{app:squeezing}, where the source Fourier variable is
$\mathbf K=\mathbf p+\mathbf q$.  Repeated tensor and derivative labels below
denote the finite contractions of that expression.  Write the kinematic
factor as
$\mathcal U^{ss'}_{ijmn}{}^{\mu\cdots\nu\cdots}
\equiv\mathcal P^{ss'}_{ijmn}{}^{\mu\cdots\nu\cdots}/
(2\sqrt{pq})$.  On the compact positive detector band
	$\mathcal B_D$ selected by the window $W_D$, whose center is denoted by
	$p_0$, define the componentwise majorant.  In the estimates below
	$p$ denotes an observed momentum inside this band, so $p\sim p_0$ and
	we use $p$ rather than $p_0$ except when identifying the window center
	\begin{equation}
		\mathcal U^{ss'}_{*,ijmn}{}^{\mu\cdots\nu\cdots}(\mathbf q)\equiv\sup_{\mathbf p\in\mathcal B_D}\left|\mathcal U^{ss'}_{ijmn}{}^{\mu\cdots\nu\cdots}(\mathbf p,\mathbf q)\right|.
	\end{equation}
Compactness of the band and the finite-order kinematic structure allow
dimensionless constants $A^{ss'}_{ijmn,\mu\cdots\nu\cdots;\beta}$ such that
$|\partial_{\mathbf p}^{\beta}\mathcal U^{ss'}_{ijmn}{}^{\mu\cdots\nu\cdots}|
\le A^{ss'}_{ijmn,\mu\cdots\nu\cdots;\beta}p^{-|\beta|}
\mathcal U^{ss'}_{*,ijmn}{}^{\mu\cdots\nu\cdots}(\mathbf q)$.
The multi-index Leibniz rule then gives
	\begin{equation}
		\begin{aligned}
			\partial_{\mathbf p}^{\alpha}
			\Xi^{ss'}_{\mathbf p,\mathbf q}(t)
			&=\sum_{\beta\le\alpha}
			\binom{\alpha}{\beta}
			\left(\partial_{\mathbf p}^{\beta}
			\mathcal U^{ss'}_{ijmn}{}^{\mu\cdots\nu\cdots}\right)\\
			&\quad\times
			\left.
			\partial_{\mathbf K}^{\alpha-\beta}
			\widetilde\xi^{ijmn}_{\mu\cdots\nu\cdots}(\mathbf K,t)
			\right|_{\mathbf K=\mathbf p+\mathbf q}.
		\end{aligned}
	\end{equation}
Combining this identity with the componentwise moment bound gives
	\begin{equation}
		\begin{aligned}
			\left|\partial_{\mathbf p}^{\alpha}
			\Xi^{ss'}_{\mathbf p,\mathbf q}(t)\right|
			&\le\sum_{\beta\le\alpha}
			\binom{\alpha}{\beta}
			A^{ss'}_{ijmn,\mu\cdots\nu\cdots;\beta}
			p^{-|\beta|}\\
			&\quad\times
			\mathcal U^{ss'}_{*,ijmn}{}^{\mu\cdots\nu\cdots}(\mathbf q)\\
			&\quad\times
			\left(R^{ijmn}_{\mu\cdots\nu\cdots;\alpha-\beta}(t)\right)^{|\alpha-\beta|}
			\xi^{ijmn}_{*,\mu\cdots\nu\cdots}(t).
		\end{aligned}
	\end{equation}
Since only finitely many tensor and derivative structures appear in a
chosen EFT operator, this componentwise bound can be summarized as
	\begin{equation}
		\left| \partial_{\mathbf p}^{\alpha}\Xi^{ss'}_{\mathbf p,\mathbf q}(t)\right| \le C_\alpha L_\Xi^{|\alpha|}\Xi_*(t,\mathbf q).
	\end{equation}
Here $C_\alpha$, $L_\Xi$, and $\Xi_*$ are positive majorants built from
the finite set of componentwise moment bounds, with $\Xi_*$ containing the
products $\mathcal U_*\xi_*$.  On the relevant source-time interval, one may choose
$L_\Xi\sim R_{\rm eff}+p^{-1}$ uniformly on the detector band, with $R_{\rm eff}$ a majorant of
the relevant componentwise source sizes.  For one external momentum
derivative this gives
	\begin{equation}
		|\nabla_{\mathbf p}\Xi^{ss'}_{\mathbf p,\mathbf q}(t)| \lesssim (R_{\rm eff}+p^{-1})\Xi_*(t,\mathbf q).
	\end{equation}

The same external-momentum smoothness is inherited by the nonlinear Volterra
numerators.  From the definitions in the main text,
	\begin{align}
		Q_{IP}(t) &= \int dJ\,\Xi_{IJ}(t)e^{iqt}\alpha^*_{JP}(t), \\
		R_{IP}(t) &= \int dJ\,\Xi_{IJ}(t)e^{iqt}\beta^*_{JP}(t)\,.
	\end{align}
When differentiating with respect to the external momentum in $I$, the
coefficients $\alpha_{JP}$ and $\beta_{JP}$ carry only the labels $J,P$.
Thus the derivative acts on $\Xi_{IJ}$, and
	\begin{equation}
		|\nabla_I^N(Q,R)_{IP}(t)| \le C_NL_\Xi^N(Q_*,R_*)_{IP}(t)\,,
	\end{equation}
with positive majorants $Q_*,R_*$ controlled by the finite-band Volterra
solution.  The same estimate holds for the temporal Fourier transforms
$(\widetilde Q,\widetilde R)(\omega)$, assuming the source time profile is
integrable on the detector band.  Squeezing can change the majorants, but it
does not create an external momentum phase whose derivatives grow with $r$.

We now define the amplitude whose smoothness is actually used in the
far-field integrals for the displayed $Q$ part.  In the angular integral of
Appendix~\ref{app:angular},
	\begin{equation}
		A_Q^{ss'}(p\hat{p},\omega,\mathbf q) = W_D(p)\mathcal C_{p\hat{p},s}\, \widetilde Q_{p\hat{p} s,\mathbf q s'}(\omega)\,.
	\end{equation}
The angular integral has the form
	\begin{equation}
		\int d\Omega_{\hat{p}}\, A_Q^{ss'}(p\hat{p},\omega,\mathbf q) e^{ipr\hat{p}\cdot\hat{n}}\,.
	\end{equation}
Angular derivatives of $A_Q$ are derivatives along the sphere
$|\mathbf p|=p$.  Each such derivative differentiates the external momentum
$\mathbf p=p\hat{p}$, and therefore brings at most a factor $pL_\Xi$,
together with fixed derivatives of $\mathcal C_{\mathbf p,s}$.  On the
detector band, with fixed channel constants included,
	\begin{equation}
		|\nabla_{S^2}^mA_Q^{ss'}| \le C_m(pL_\Xi)^m A_Q^{{\rm maj},ss'}(\omega,\mathbf q).
	\end{equation}
Here $m$ is a positive integer, and the estimate holds pointwise on the
detector band.  The detector window $W_D(p)$ depends only on the radial
momentum, so angular derivatives do not act on it.  The detector bandwidth
therefore does not enter this angular estimate.

In the radial pole extraction of Appendix~\ref{app:retarded-pole} the endpoint angular expansion
has already been taken.  The radial amplitudes are
	\begin{align}
		a_{Q,+}^{ss'}(p,\omega,\mathbf q) \nonumber &= \frac{2\pi p}{ir}\, W_D(p)\mathcal C_{p\hat{n},s}\, \widetilde Q_{p\hat{n} s,\mathbf q s'}(\omega), \\
		a_{Q,-}^{ss'}(p,\omega,\mathbf q) &= -\frac{2\pi p}{ir}\, W_D(p)\mathcal C_{-p\hat{n},s}\, \widetilde Q_{-p\hat{n} s,\mathbf q s'}(\omega)\,.
	\end{align}
For fixed $J=(\mathbf q,s')$, they appear in integrals of the form
	\begin{equation}
		I_\pm= \sum_s\int_{\mathbb R}dp\, a_{Q,\pm}^{ss'}(p,\omega,\mathbf q) \frac{e^{\pm ipr}}{\omega-p+i0}\,,
	\end{equation}
where the half-line $p>0$ has been extended to the real line because
$W_D$ is compactly supported inside a positive-frequency band.  The product
rule, the bounds on $\widetilde Q$, and the window derivative bounds give
	\begin{equation}
		L_D\sim R_{\rm eff}+p^{-1}+\omega_D^{-1},
	\end{equation}
and hence
	\begin{equation}
		\left| \partial_p^N a_{Q,\pm}^{ss'}(p,\omega,\mathbf q) \right| \le C_NL_D^N a_Q^{{\rm maj},ss'}(\omega,\mathbf q)\,.
	\end{equation}
The harmless factor $1/r$ in $a_{Q,\pm}$ is included in the majorant
$a_Q^{\rm maj}$.  Crucially, no derivative of $a_{Q,\pm}$ produces a positive
power of $r$; all $r$-oscillation in the radial integral is the explicit
propagation phase $e^{\pm ipr}$.

\section{Angular Stationary Phase}
\label{app:angular}

This appendix derives the angular endpoint expansion used before the radial
pole extraction.  Let the observation point be $\mathbf x=r\hat{n}$, and
choose the polar axis of $\hat{p}$ along $\hat{n}$.  Write
	\begin{equation}
		\mu=\cos\theta, \qquad d\Omega_{\hat{p}}=d\mu\,d\varphi \,.
	\end{equation}
For fixed $p,\omega,\mathbf q$, let $A(p\hat{p},\omega,\mathbf q)$ denote the smooth amplitude $A_Q^{ss'}$ defined in Appendix~\ref{app:locality}, and define
its azimuthal average by
	\begin{equation}
		\overline A(p,\mu,\omega,\mathbf q) = \int_0^{2\pi}d\varphi\, A(p,\mu,\varphi,\omega,\mathbf q)\,.
	\end{equation}
The angular integral is then
	\begin{equation}
		J(p,r) = \int_{-1}^{1}d\mu\, \overline A(p,\mu,\omega,\mathbf q)e^{ipr\mu}\,.
	\end{equation}
Although the stationary directions of the original spherical phase are
$\hat{p}=\pm\hat{n}$, in the variable $\mu$ they are simply the two endpoints
$\mu=\pm1$.

One integration by parts gives
	\begin{align}
		J(p,r) \nonumber &= \frac{1}{ipr} \left[ e^{ipr}\overline A(p,1,\omega,\mathbf q) -e^{-ipr}\overline A(p,-1,\omega,\mathbf q) \right] \\
		&\quad -\frac{1}{ipr} \int_{-1}^{1}d\mu\, \partial_\mu\overline A(p,\mu,\omega,\mathbf q)e^{ipr\mu}\,.
	\end{align}
At the two poles the azimuthal coordinate is degenerate, so smoothness on the
sphere implies
	\begin{align}
		\overline A(p,1,\omega,\mathbf q) &=2\pi A(p\hat{n},\omega,\mathbf q), \\
		\overline A(p,-1,\omega,\mathbf q) &=2\pi A(-p\hat{n},\omega,\mathbf q)\,.
	\end{align}
Therefore
	\begin{align}
		J(p,r) \nonumber &= \frac{2\pi}{ipr} \left[ e^{ipr}A(p\hat{n},\omega,\mathbf q) -e^{-ipr}A(-p\hat{n},\omega,\mathbf q) \right] \\
		&\quad +I_{\rm rem}(p,r)\,,
	\end{align}
with
	\begin{equation}
		I_{\rm rem}(p,r) = -\frac{1}{ipr} \int_{-1}^{1}d\mu\, \partial_\mu\overline A(p,\mu,\omega,\mathbf q)e^{ipr\mu}\,.
	\end{equation}

The derivative $\partial_\mu\overline A$ is not singular at
$\mu=\pm1$.  Since $\mu=\cos\theta$,
	\begin{equation}
		\partial_\mu=-\frac{1}{\sin\theta}\partial_\theta\,,
	\end{equation}
which appears singular at the poles.  The point is that this derivative acts
on the azimuthal average, not on a single coordinate representative of the
spherical function.  Near the north pole,
	\begin{equation}
		\partial_\mu\overline A(p,\mu,\omega,\mathbf q) = -\frac{1}{\sin\theta} \int_0^{2\pi}d\varphi\, \partial_\theta A(p,\theta,\varphi,\omega,\mathbf q)\,.
	\end{equation}
We therefore define
	\begin{equation}
		F_N(\theta) = \int_0^{2\pi}d\varphi\, \partial_\theta A(p,\theta,\varphi,\omega,\mathbf q)\,.
	\end{equation}
This definition isolates the numerator of the apparently singular ratio
$-F_N(\theta)/\sin\theta$.  To prove regularity it is enough to show that
this numerator vanishes at least linearly as $\theta\to0$.  Indeed,
	\begin{equation}
		\partial_\mu\overline A=-\frac{F_N(\theta)}{\sin\theta}\,.
	\end{equation}
At the north pole the direction $\varphi$ is only a coordinate label.  For a
smooth function on the sphere,
$\partial_\theta A(p,0,\varphi,\omega,\mathbf q)$ is the directional
derivative of $A$ at $p\hat{n}$ along the unit tangent
$\mathbf e_\theta(\varphi)$.  Hence
	\begin{equation}
		F_N(0) = \int_0^{2\pi}d\varphi\, \nabla_{S^2}A(p\hat{n})\cdot\mathbf e_\theta(\varphi) =0\,,
	\end{equation}
because the average of the unit tangent vector
$\mathbf e_\theta(\varphi)$ over the azimuthal angle vanishes.  Thus the
numerator has a zero at the pole.

It remains to control how fast $F_N$ can grow away from zero.  Along each
fixed-$\varphi$ meridian, $\theta$ is geodesic distance on the unit sphere,
and the meridian tangent is parallel along that geodesic.  Therefore
	\begin{equation}
		\frac{d}{d\theta}\partial_\theta A = \nabla_{S^2}^2A(\mathbf e_\theta,\mathbf e_\theta)\,,
	\end{equation}
while Appendix~\ref{app:locality} gives the pointwise angular-smoothness estimate
	\begin{equation}
		|\nabla_{S^2}^2A| \le C(pL_\Xi)^2A_* \,.
	\end{equation}
Using $F_N(0)=0$, this gives
	\begin{equation}
		|F_N(\theta)| \le C|\theta|\,(pL_\Xi)^2A_* \,.
	\end{equation}
Since $\sin\theta\sim\theta$ near the north pole,
	\begin{equation}
		|\partial_\mu\overline A| \le C(pL_\Xi)^2A_*
	\end{equation}
there.  The south pole is identical after replacing $\theta$ by
$\pi-\theta$ and defining the corresponding numerator $F_S$.  Thus the
factor $1/\sin\theta$ is only a coordinate singularity: the azimuthal
average supplies a numerator that vanishes at the same rate, and
$\partial_\mu\overline A$ is bounded on the closed interval $[-1,1]$.

The next term in the endpoint expansion follows by integrating $I_{\rm rem}$ once more by parts:
	\begin{align}
		I_{\rm rem}(p,r) \nonumber &= -\frac{1}{(ipr)^2} \left[ e^{ipr}\partial_\mu\overline A(p,1) -e^{-ipr}\partial_\mu\overline A(p,-1) \right] \\
		&\quad +\frac{1}{(ipr)^2} \int_{-1}^{1}d\mu\, \partial_\mu^2\overline A(p,\mu,\omega,\mathbf q)e^{ipr\mu}\,.
	\end{align}
The first line is the first angular correction.  From
Appendix~\ref{app:locality},
	\begin{equation}
		|\nabla_{S^2}^2A| \le C(pL_\Xi)^2A_* \,,
	\end{equation}
where $A_*$ is a band-limited majorant.  Therefore
	\begin{equation}
		|\partial_\mu\overline A(p,\pm1)| \le C(pL_\Xi)^2A_* \,,
	\end{equation}
and the first correction satisfies
	\begin{equation}
		|I_{\rm rem}^{(1)}(p,r)| \le \frac{C(pL_\Xi)^2A_*}{(pr)^2}\,.
	\end{equation}
The leading endpoint term has size $A_*/(pr)$.  Thus the relative size of
the first angular correction is bounded by
	\begin{equation}
		\frac{|I_{\rm rem}^{(1)}|}{|J_{\rm lead}|} \lesssim \frac{pL_\Xi^2}{r}
	\end{equation}
on the detector band.  Since $L_D\ge L_\Xi$, a simple
sufficient condition for keeping only the leading angular endpoints is the
Fraunhofer condition
	\begin{equation}
		r\gg pL_D^2 \,.
	\end{equation}
If this condition is not imposed, the endpoint expansion remains valid, but
the subleading endpoint derivatives should be retained.  The angular analysis
by itself only separates the two geometric propagation directions,
$\hat{p}=\hat{n}$ and $\hat{p}=-\hat{n}$.  The absence of an advanced
contribution is established only after the radial Plemelj analysis in
Appendix~\ref{app:retarded-pole}.

\section{Retarded Pole and Absence of Advanced Wave}
\label{app:retarded-pole}

This appendix performs the radial pole extraction.  After the angular endpoint
expansion, each source-dependent term has a radial integral of the form
	\begin{equation}
		I_\pm = \int_{\mathbb R}dp\, a_\pm(p,\omega,\mathbf q) \frac{e^{\pm ipr}}{\omega-p+i0}\,,
	\end{equation}
where, for the displayed $Q$-part, $a_\pm(p,\omega,\mathbf q)\equiv
\sum_s a_{Q,\pm}^{ss'}(p,\omega,\mathbf q)$; the $R$-part is analogous.
The amplitudes $a_{Q,\pm}^{ss'}$ are defined in
Appendix~\ref{app:locality}.  The half-line $p>0$
has been extended to
$\mathbb R$ because the detector window is supported inside a positive
frequency band.

The distributional identity
	\begin{equation}
		\frac{1}{\omega-p+i0} = {\rm PV}\frac{1}{\omega-p} -i\pi\delta(\omega-p)
	\end{equation}
fixes the retarded boundary value.  For the outgoing branch,
	\begin{equation}
		I_+ = \int_{\mathbb R}dp\, a_+(p,\omega,\mathbf q) \frac{e^{ipr}}{\omega-p+i0}\,.
	\end{equation}
Assume first that $\omega$ lies inside the detector band.  Set
$p=\omega+u$, and choose a smooth cutoff $\eta(u)$ which equals one near
$u=0$.  The principal-value part contains
	\begin{equation}
		-e^{i\omega r} {\rm PV}\int du\, \frac{a_+(\omega+u,\omega,\mathbf q)e^{iur}}{u}\,.
	\end{equation}
Split
	\begin{equation}
		a_+(\omega+u) = a_+(\omega)\eta(u) + \left[a_+(\omega+u)-a_+(\omega)\eta(u)\right]\,.
	\end{equation}
The bracket vanishes at $u=0$, so
	\begin{equation}
		g_+(u;\omega) = \frac{a_+(\omega+u)-a_+(\omega)\eta(u)}{u}
	\end{equation}
is smooth and compactly supported.  The singular piece gives
	\begin{equation}
		{\rm PV}\int du\,\frac{\eta(u)e^{iur}}{u} = i\pi+S_\eta(r)\,,
	\end{equation}
where $S_\eta(r)$ is rapidly decreasing. To justify this identity, define
$I_\eta(r)={\rm PV}\int du\,\eta(u)e^{iur}/u$.  Differentiating with respect
to $r$ removes the principal-value singularity and gives
\begin{equation}
\frac{d I_\eta}{dr}
=
i\int du\,\eta(u)e^{iur}\,.
\end{equation}
Since $\eta$ is smooth and compactly supported, the right-hand side is rapidly
decreasing as $r\to+\infty$.  Hence $I_\eta(r)$ approaches a constant up to a
remainder $S_\eta(r)=\mathcal{O}(r^{-N})$ for any $N$.  The constant is fixed by the
standard distributional identity
\begin{equation}
\int_0^\infty ds\,e^{isu}
=
\pi\delta(u)+i\,{\rm PV}\frac{1}{u}\,.
\end{equation}
Using $\eta(0)=1$, one obtains
\begin{equation}
	\lim_{r\to+\infty}I_\eta(r)=i\pi\,,
\end{equation}
which exactly gives the identity above. Hence the local singular part of
the principal value is $-i\pi a_+(\omega)e^{i\omega r}$, up to a smooth
remainder.  The delta term contributes another
$-i\pi a_+(\omega)e^{i\omega r}$.  Therefore
	\begin{equation}
		I_+ = -2\pi i\,a_+(\omega,\omega,\mathbf q)e^{i\omega r} +R_+(r;\omega,\mathbf q)\,.
	\end{equation}
The remainder $R_+$ is an oscillatory integral with no local pole.  Its
physical meaning is that nearby radial momenta carry rapidly varying phases
$e^{ipr}$; once the pole part is removed, neighboring points in the integral
destructively interfere.  We make this cancellation quantitative below by
using the smoothness estimates of Appendix~\ref{app:locality}.
In the field operator this radial factor is multiplied by the explicit time
phase $e^{-i\omega t}$, so the outgoing pole gives
	\begin{equation}
		e^{-i\omega t}e^{i\omega r} = e^{-i\omega(t-r)}\,.
	\end{equation}
This is the retarded propagation phase.

For the incoming branch,
	\begin{equation}
		I_- = \int_{\mathbb R}dp\, a_-(p,\omega,\mathbf q) \frac{e^{-ipr}}{\omega-p+i0}\,.
	\end{equation}
The same local calculation gives the principal-value singular term
$+i\pi a_-(\omega)e^{-i\omega r}$, whereas the delta term gives
$-i\pi a_-(\omega)e^{-i\omega r}$.  These two terms cancel exactly:
	\begin{equation}
		+i\pi a_-(\omega)e^{-i\omega r} -i\pi a_-(\omega)e^{-i\omega r} =0\,.
	\end{equation}
Thus the incoming geometric endpoint has no local on-shell term proportional
to $e^{-i\omega(t+r)}$.  It leaves only a smooth oscillatory remainder.  If
$\omega$ lies outside the detector band, the denominator has no zero on the
support of $a_\pm$, and both $I_+$ and $I_-$ are smooth oscillatory
integrals from the start.

The remainders are controlled by repeated integration by parts.  If $g(p)$ is a smooth compactly supported function, then
	\begin{equation}
		\left| \int_{\mathbb R}dp\,g(p)e^{\pm ipr} \right| \le \frac{\|\partial_p^Ng\|_{L^1}}{r^N}\,.
	\end{equation}
After subtracting the local pole, the functions $g_\pm$ and the cutoff
remainders are smooth compactly supported amplitudes.
Appendix~\ref{app:locality} gives
	\begin{equation}
		|\partial_p^N a_\pm(p,\omega,\mathbf q)| \le C_NL_D^N a_*(\omega,\mathbf q)\,,
	\end{equation}
and therefore, for every fixed $N$ allowed by the source and detector
smoothness,
	\begin{equation}
		|R_\pm(r;\omega,\mathbf q)| \le C_Na_*(\omega,\mathbf q) \left(\frac{L_D}{r}\right)^N \,.
	\end{equation}
Thus all non-pole radial pieces are suppressed in the far-field regime
$r\gg L_D$, while the only local on-shell pole is the outgoing retarded one.

Substituting the outgoing amplitudes gives the source part of the finite-band
detector field.  Define
	\begin{align}
		\mathcal V_s(\omega;x) \nonumber &= -2\pi i e^{-i\omega(t-r)} \frac{2\pi\omega}{ir} W_D(\omega)\mathcal C_{\omega\hat{n},s}\,.
	\end{align}
Then
	\begin{align}
		\hat B_{{\rm src},D}^{(+)}(x) \nonumber &= \sum_{ss'}\int d^3q\int\frac{d\omega}{2\pi}\, \mathcal V_s(\omega;x) \\
		\times& \bigl[ \widetilde R_{\omega\hat{n} s,\mathbf q s'}(\omega)\hat a_{\mathbf q,s'} +\widetilde Q_{\omega\hat{n} s,\mathbf q s'}(\omega) \hat a^\dagger_{\mathbf q,s'} \bigr] \\
		+&\mathcal{O}\left[\left(\frac{L_D}{r}\right)^N\right]\,.
	\end{align}
Equivalently, in the compact notation of the main text, the pole extraction
justifies writing the local far field in terms of the outgoing operator
$\hat b_{\omega\hat{n},s}^{\rm out}$.

Finally we record the relation between the on-shell Volterra numerators and
the Bogoliubov out coefficients.  If the source interaction has a future
out-region and the full integrals exist, then
	\begin{align}
		\bigl(\beta^{\rm out},\alpha_{\rm sc}^{\rm out}\bigr)_{IP} \nonumber &= -i\int_{-\infty}^{+\infty}dt\, e^{ip t} \bigl(Q,R\bigr)_{IP}(t) \\
		&= -i\bigl(\widetilde Q,\widetilde R\bigr)_{IP}(p)\,.
	\end{align}
The radial pole extraction is the step that sets the external line on shell,
$\omega=p$, so this is precisely the relation used in the main text.
Physically, this means that a far-field observer measures the asymptotic
scattering output of the source.

\section{Same-Cone Suppression}
\label{app:same-cone}

This appendix proves the same-cone estimate used in the main text.  We work
in the leading-Magnus approximation of Appendix~\ref{app:squeezing}, and write
$\mathcal G=\mathcal G(+\infty)$.  Polarization labels are kept implicit
inside fixed finite constants.  On the detector band we assume the
back-to-back envelope
	\begin{equation}
		{|\mathcal G_{\mathbf p,\mathbf q}^{ss'}| \le G_0 \exp\left(-c\frac{|\mathbf p+\mathbf q|}{k_*}\right)}, \qquad {c}>0\,.
	\end{equation}
Here $k_*$ is the total-momentum width of the pair-production kernel, and
the non-relativistic same-cone regime is {$k_*\ll p+q$}.
The constants below also absorb the finite detector bandwidth and the chosen
normalization of the momentum measure.

We first record the elementary convolution estimate.  Define
	\begin{equation}
		{h_\lambda(\mathbf{x}) = \exp\left(-\lambda\frac{|\mathbf{x}|}{k_*}\right).}
	\end{equation}
{For any $0<d<c$},
	\begin{align}
		&{(h_c*h_c)(\mathbf{x}) \le L_{c,d}h_{d}(\mathbf{x}),} \\
		&{L_{c,d}= \int d^3u\, \exp\left[-(c-d)\frac{|\mathbf u|}{k_*}\right] = \frac{8\pi k_*^3}{(c-d)^3}.}
	\end{align}
Indeed,
	\begin{equation}
		{c|\mathbf{x}-\mathbf u|+c|\mathbf u| \ge d|\mathbf{x}|+(c-d)|\mathbf u|,}
	\end{equation}
so the integral over $\mathbf u$ is bounded by the expression above.  The
same argument gives, by induction,
	\begin{equation}
		{h_c^{*n}(\mathbf{x}) \le L_{c,d}^{\,n-1}h_{d}(\mathbf{x}), \qquad n\ge1 .}
	\end{equation}

The leading-Magnus coefficients have the series
	\begin{align}
		\beta^{\rm out} &= \sum_{m=0}^{\infty} \frac{\mathcal G(\mathcal G^\dagger\mathcal G)^m}{(2m+1)!}, \\
		\alpha_{\rm sc}^{\rm out} &= \alpha^{\rm out}-I = \sum_{m=1}^{\infty} \frac{(\mathcal G\mathcal G^\dagger)^m}{(2m)!}\,.
	\end{align}
Each odd chain in $\beta^{\rm out}$ is a convolution in the
anti-diagonal variable {$\mathbf p+\mathbf q$}.  Applying the bound above to
the $2m+1$ kernels gives
	\begin{equation}
		{\left| \bigl[\mathcal G(\mathcal G^\dagger\mathcal G)^m\bigr]_{\mathbf p,\mathbf q} \right| \le G_0^{2m+1}L_{c,d}^{2m} \exp\left(-d\frac{|\mathbf p+\mathbf q|}{k_*}\right)}\,.
	\end{equation}
After summing the odd series,
	\begin{equation}
		{|\beta^{\rm out}_{\mathbf p,\mathbf q}| \le B_{d} \exp\left(-d\frac{|\mathbf p+\mathbf q|}{k_*}\right)}, \quad {B_{d} = \frac{\sinh(G_0L_{c,d})}{L_{c,d}}} \,.
	\end{equation}
Thus $\beta^{\rm out}$ is localized near the anti-diagonal
{$\mathbf q\simeq-\mathbf p$}.

For the even chains, the first nontrivial factor is
	\begin{equation}
		{(\mathcal G\mathcal G^\dagger)_{\mathbf p,\mathbf q} = \sum_\lambda\int d^3\ell\, \mathcal G_{\mathbf p,\boldsymbol\ell}^{s\lambda} \mathcal G_{\mathbf q,\boldsymbol\ell}^{s'\lambda *}.}
	\end{equation}
The relevant variable is now {$\mathbf p-\mathbf q$}, because
	\begin{equation}
		{|\mathbf p-\mathbf q| \le |\mathbf p+\boldsymbol\ell|+|\mathbf q+\boldsymbol\ell|.}
	\end{equation}
The same convolution estimate therefore gives, for $m\ge1$,
	\begin{equation}
		{\left| \bigl[(\mathcal G\mathcal G^\dagger)^m\bigr]_{\mathbf p,\mathbf q} \right| \le G_0^{2m}L_{c,d}^{2m-1} \exp\left(-d\frac{|\mathbf p-\mathbf q|}{k_*}\right)}\,.
	\end{equation}
Summing the even series yields
	\begin{align}
		{|\alpha^{\rm out}_{{\rm sc},\mathbf p,\mathbf q}|} &\le {A_{d} \exp\left(-d\frac{|\mathbf p-\mathbf q|}{k_*}\right)}, \\
		{A_{d}} &{= \frac{\cosh(G_0L_{c,d})-1}{L_{c,d}}} \,.
	\end{align}
Hence $\alpha_{\rm sc}^{\rm out}$ is forward-localized near
{$\mathbf q\simeq\mathbf p$}, while $\beta^{\rm out}$ is anti-diagonal.

We now allow a small angular separation between the two far-field detector
points.  For the zero-delay spatial correlation take
	\begin{equation}
		x_1=(t,r\hat{n}_1), \quad x_2=(t,r\hat{n}_2), \quad \hat{n}_1\cdot\hat{n}_2=\cos\delta\theta \,,
	\end{equation}
with $\delta\theta\ll1$.  Both points have the same retarded time
$\tau=t-r$.  The detector kernels are
	\begin{equation}
		\mathcal U_{i,s}({p_i}) = \frac{2\pi{p_i}}{ir} W_D({p_i})\mathcal C_{{p_i}\hat{n}_i,s} e^{-i{p_i}\tau}, \qquad i=1,2\,.
	\end{equation}
The field-level normal and anomalous kernels are
	\begin{align}
		N_D(x_1,x_2) &= \sum_{ss'}\int_0^\infty d{p_1}\,d{p_2}\, \nonumber\\
		&\quad\times \mathcal U^*_{1,s}({p_1}) \mathcal U_{2,s'}({p_2}) n^{\rm out}_{12;ss'}({p_1},{p_2}), \\
		M_D(x_1,x_2) &= \sum_{ss'}\int_0^\infty d{p_1}\,d{p_2}\, \nonumber\\
		&\quad\times \mathcal U_{1,s}({p_1}) \mathcal U_{2,s'}({p_2}) m^{\rm out}_{12;ss'}({p_1},{p_2}) \,.
	\end{align}
Here the mode-level Bogoliubov overlaps are, with
$\mathbf p_1={p_1}\hat{n}_1$ and
$\mathbf p_2={p_2}\hat{n}_2$,
	\begin{align}
		n^{\rm out}_{12;ss'}({p_1},{p_2}) &= \sum_\lambda\int d^3q\, \beta^{{\rm out}*}_{{\mathbf p_1}s,\mathbf q\lambda} \beta^{\rm out}_{{\mathbf p_2}s',\mathbf q\lambda}, \\
		m^{\rm out}_{12;ss'}({p_1},{p_2}) &= \sum_\lambda\int d^3q\, \alpha^{\rm out}_{{\mathbf p_1}s,\mathbf q\lambda} \beta^{\rm out}_{{\mathbf p_2}s',\mathbf q\lambda}\,.
	\end{align}
The Gaussian identity therefore gives
	\begin{align}
		\gamma_N(x_1,x_2) &= \frac{N_D(x_1,x_2)} {\sqrt{N_D(x_1,x_1)N_D(x_2,x_2)}} , \\
		\gamma_M(x_1,x_2) &= \frac{M_D(x_1,x_2)} {\sqrt{N_D(x_1,x_1)N_D(x_2,x_2)}} , \\
		g^{(2)}(x_1,x_2) &= 1+|\gamma_N(x_1,x_2)|^2+|\gamma_M(x_1,x_2)|^2 \,.
	\end{align}
Only in the strictly coincident limit does the normal term become unity and
recover the expression $2+|M_D|^2/N_D^2$.

The normal kernel contains the overlap of two anti-diagonal $\beta$
envelopes.  {For the overlap integrals below we choose a further weakened
exponent $0<d'<d$.}  The inequality
	\begin{equation}
		{|\mathbf p_1-\mathbf p_2| \le |\mathbf p_1+\mathbf q|+|\mathbf p_2+\mathbf q|}
	\end{equation}
gives
	\begin{equation}
		|n^{\rm out}_{12;ss'}({p_1},{p_2})| \le {C_N(d,d') \exp\left(-d'\frac{D_-(p_1,p_2,\delta\theta)}{k_*}\right)}\,,
	\end{equation}
	with
	\begin{equation}
		{C_N(d,d')=N_{\rm pol}B_d^2V_{d,d'}\,,\qquad N_{\rm pol}=2}\,,
	\end{equation}
where $V_{d,d'}$ is defined below and $N_{\rm pol}$ counts the two tensor
polarizations.  The momentum-measure convention is already included in
$B_d$ and $V_{d,d'}$.  The momentum cost is
	\begin{equation}
		\begin{aligned}
		D_-({p_1},{p_2},\delta\theta) &= |{p_1}\hat{n}_1-{p_2}\hat{n}_2| \\
		&= \sqrt{{p_1}^2+{p_2}^2-2{p_1}{p_2}\cos\delta\theta}\,.
		\end{aligned}
	\end{equation}
Thus the normal correlation is forward in the two detected directions.  In
the strictly collinear case $D_-=|{p_1}-{p_2}|$.  For small angular
separation,
	\begin{equation}
		D_-^2 = ({p_1}-{p_2})^2+{p_1}{p_2}\delta\theta^2 +{\mathcal{O}}(\delta\theta^4)\,.
	\end{equation}
In a narrow band centered at ${p_0}$, this becomes
	\begin{equation}
		D_-({p_0},{p_0},\delta\theta) = 2{p_0}\sin\frac{\delta\theta}{2} \simeq {p_0}\delta\theta \,.
	\end{equation}
Therefore the normal HBT term remains close to its coincident value only
inside the angular coherence scale $\delta\theta\lesssim k_*/{p_0}$.

The anomalous kernel contains the overlap of a forward
$\alpha_{\rm sc}$ envelope with an anti-diagonal $\beta$ envelope, plus
the identity part of $\alpha^{\rm out}=I+\alpha_{\rm sc}^{\rm out}$.  The
identity contribution is
	\begin{equation}
		m^{\rm out}_{I;ss'}({p_1},{p_2}) = {\beta^{\rm out}_{\mathbf p_2s',\mathbf p_1s}}\,,
	\end{equation}
and obeys
	\begin{equation}
		|m^{\rm out}_{I;ss'}({p_1},{p_2})| \le {B_{d} \exp\left(-d'\frac{|\mathbf p_1+\mathbf p_2|}{k_*}\right)}\,,
	\end{equation}
For the scattered part, the triangle inequality
	\begin{equation}
		{|\mathbf p_1+\mathbf p_2| \le |\mathbf p_1-\mathbf q|+|\mathbf p_2+\mathbf q|}
	\end{equation}
gives
	\begin{equation}
		|m^{\rm out}_{{\rm sc};ss'}({p_1},{p_2})| \le {A_{d}B_{d}V_{d,d'} \exp\left(-d'\frac{D_+(p_1,p_2,\delta\theta)}{k_*}\right)}\,,
	\end{equation}
where
	\begin{equation}
		{V_{d,d'} = \int d^3u\, \exp\left[-(d-d')\frac{|\mathbf u|}{k_*}\right] = \frac{8\pi k_*^3}{(d-d')^3}}\,,
	\end{equation}
and
	\begin{equation}
		\begin{aligned}
		D_+({p_1},{p_2},\delta\theta) &= |{p_1}\hat{n}_1+{p_2}\hat{n}_2| \\
		&= \sqrt{{p_1}^2+{p_2}^2+2{p_1}{p_2}\cos\delta\theta}\,.
		\end{aligned}
	\end{equation}
Combining the identity and scattered pieces gives the same-cone envelope
	\begin{equation}
		|m^{\rm out}_{12;ss'}({p_1},{p_2})| \le {C_{\rm sq}(d,d') \exp\left(-d'\frac{D_+(p_1,p_2,\delta\theta)}{k_*}\right)}\,,
	\end{equation}
with, for example,
	\begin{equation}
		{C_{\rm sq}(d,d') = B_{d}+A_{d}B_{d}V_{d,d'}} \,.
	\end{equation}
In the strictly collinear same-cone case $D_+={p_1}+{p_2}$.  The first
angular correction is quadratic:
	\begin{equation}
		D_+ = ({p_1}+{p_2}) - \frac{{p_1}{p_2}}{2({p_1}+{p_2})}\delta\theta^2 +{\mathcal{O}}(\delta\theta^4)\,.
	\end{equation}
For ${p_1}\simeq{p_2}\simeq{p_0}$,
	\begin{equation}
		D_+({p_0},{p_0},\delta\theta) = 2{p_0}\cos\frac{\delta\theta}{2} \simeq 2{p_0} \left(1-\frac{\delta\theta^2}{8}\right)\,.
	\end{equation}
Thus a small angular separation changes the same-cone suppression only at
order $\delta\theta^2$.

It remains to pass from the mode envelope to the finite detector channel.
For the compact detector band $\mathcal B_D$, define
	\begin{align}
		\Epsilon_D(\delta\theta)
		&\equiv \inf_{p_1,p_2\in\mathcal B_D}
		D_+(p_1,p_2,\delta\theta),\\
		C_D(x_1,x_2)
		&\equiv \sum_{ss'}\int_{\mathcal B_D}dp_1\,dp_2\,
		|\mathcal U_{1,s}(p_1)\mathcal U_{2,s'}(p_2)|\,.
	\end{align}
Both quantities are finite for a smooth compactly supported detector window.
The uniform finite-band bound is then
	\begin{equation}
		|M_D(x_1,x_2)| \le {C_D(x_1,x_2) C_{\rm sq}(d,d')
		\exp\left(-d'\frac{\Epsilon_D(\delta\theta)}{k_*}\right)}\,.
	\end{equation}
For the order-of-magnitude estimate used in the Letter, $\Epsilon$ denotes the
characteristic value of $D_+$ over the frequencies that dominate $W_D$.  In a
narrow same-cone band centered at ${p_0}$,
	\begin{equation}
		\Epsilon(\delta\theta) \sim D_+({p_0},{p_0},\delta\theta) = 2{p_0}\cos\frac{\delta\theta}{2}\,.
	\end{equation}
Therefore a small angular displacement changes the energy cost by the same
amount as $D_+$:
	\begin{equation}
		\frac{\Epsilon(\delta\theta)-\Epsilon(0)}{\Epsilon(0)} \simeq -\frac{\delta\theta^2}{8}\,.
	\end{equation}
Using the uniform bound above gives
	\begin{align}
		g^{(2)}(x_1,x_2) &\le 1+|\gamma_N(x_1,x_2)|^2 \nonumber\\
		&\quad+ {\frac{C_D^2(x_1,x_2)C_{\rm sq}^2(d,d')} {N_D(x_1,x_1)N_D(x_2,x_2)}} \nonumber\\
		&\qquad\times {\exp\left(-2d'\frac{\Epsilon_D(\delta\theta)}{k_*}\right)}\,,
	\end{align}

	To keep the effective occupation dimensionless, define
	\begin{equation}
		n_D\equiv\frac{N_D}{\mathcal N_D}=\sinh^2r_{\rm eff}\,,
		\qquad m_D\equiv\frac{M_D}{\mathcal N_D}\,,
	\end{equation}
	where $\mathcal N_D$ is fixed by the normalization of the detector
	wave-packet mode.  Hence $|M_D|/N_D=|m_D|/n_D$, and the normalization
	cancels from $g^{(2)}$.

The compact estimate quoted in the main text is the coincident or
same-coherence-cell limit, where $|\gamma_N|\simeq1$.  {This
order-of-magnitude estimate uses the inherited $\beta$-envelope scale $d$,
rather than the extra weakened exponent $d'$ used in the preceding rigorous
overlap bound.}  Applying the normalized occupation
$n_D=\sinh^2r_{\rm eff}$ defined above, the
identity part scales like $|\beta|$, while the rescattered part scales like
$|\alpha_{\rm sc}\beta|$.  Using
\[
\frac{|\alpha_{\rm sc}|}{|\beta|}
\sim
\frac{\cosh r_{\rm eff}-1}{\sinh r_{\rm eff}}
=
\tanh\frac{r_{\rm eff}}{2},
\]
one obtains
	\begin{equation}
		\frac{|M_D|}{N_D} \lesssim {\exp\left(-d\frac{\Epsilon(\delta\theta)}{k_*}\right)} \left[ \frac{1}{\sinh r_{\rm eff}} + \tanh\frac{r_{\rm eff}}{2} \right]\,,
	\end{equation}
and therefore
	\begin{align}
		&g^{(2)}(x_1,x_2)-1-|\gamma_N(x_1,x_2)|^2 \nonumber\\
		&\quad\lesssim {\exp\left(-2d\frac{\Epsilon(\delta\theta)}{k_*}\right)} \left[ \frac{1}{\sinh r_{\rm eff}} + \tanh\frac{r_{\rm eff}}{2} \right]^2 \,.
	\end{align}
For $\delta\theta=0$ and $|\gamma_N|=1$ this reduces to the same-point
formula in the main text.  {The main text therefore writes the coincident
suppression schematically as $\exp(-2d\Epsilon/k_*)$, while the preceding
two-point overlap bound may be stated with any stricter $0<d'<d$.}

\end{document}